\documentstyle[12pt,aaspp4]{article}
\begin{document}

\newcommand{\up}[1]{\ifmmode^{\rm #1}\else$^{\rm #1}$\fi}
\newcommand{\zdot}{\makebox[0pt][l]{.}}
\newcommand{\upd}{\up{d}}
\newcommand{\uph}{\up{h}}
\newcommand{\upm}{\up{m}}
\newcommand{\ups}{\up{s}}
\newcommand{\arcd}{\ifmmode^{\circ}\else$^{\circ}$\fi}
\newcommand{\arcm}{\ifmmode{'}\else$'$\fi}
\newcommand{\arcs}{\ifmmode{''}\else$''$\fi}

\title{The ARAUCARIA project: Deep near-infrared survey 
of nearby galaxies. I. The distance to the Large Magellanic Cloud 
from K-band photometry of red clump stars
\footnote{Based on  observations obtained with the NTT
telescope at the European Southern Observatory as part of 
project number 69.D-0352.
}}

\author{G. Pietrzy{\'n}ski}
\affil{Universidad de Concepci{\'o}n, Departamento de Fisica, Casilla 160--C, 
Concepci{\'o}n, Chile}
\affil{Warsaw University Observatory, Al. Ujazdowskie 4,00-478, Warsaw, Poland}
\authoremail{pietrzyn@hubble.cfm.udec.cl}
\author{W. Gieren}
\affil{Universidad de Concepci{\'o}n, Departamento de Fisica, Casilla 160--C, 
Concepci{\'o}n, Chile}
\authoremail{wgieren@coma.cfm.udec.cl}

\begin{abstract}
We have obtained deep imaging in the near-infrared J and K bands for 2 nearby 
fields in
the bar of the LMC with the ESO NTT telescope, under exquisite seeing
conditions. The K, J-K color-magnitude diagrams constructed from these
data are of outstanding photometric quality and reveal the presence of several
hundreds of red clump stars.

Using the calibration of Alves for the K-band absolute 
magnitude of {\it Hipparcos}-observed red clump stars in the solar neigbourhood  
we derive a distance 
modulus to our observed LMC fields of 18.487 mag. Applying a correction for the 
tilt
of the LMC bar with respect to the line of sight according to the geometrical 
model
of van der Marel et al., the corresponding LMC barycenter distance is 18.501 
mag.
The random error on this result is +/- 0.008 
mag,
whereas the systematic uncertainty on this distance result due to the 
photometric
zero point uncertainty in our LMC K-band photometry, to the uncertainty of the
{\it Hipparcos}-based absolute magnitude calibration of red clump stars in the 
solar
neighborhood, and due to reddening uncertainties mounts up to +/- 0.048 mag.

If we adopt a K-band population correction of -0.03 mag, as done by Alves et al.
2002, to account for the difference in age and metallicity between the solar
neighborhood and LMC red clump star populations, we obtain an LMC barycenter
distance modulus of 18.471 mag from our data. This is in excellent agreement
with the result of Alves et al., and of another very recent study of Sarajedini 
et al. (2002)
obtained from K-band photometry. However, we emphasize that current
model predictions about the uncertainties of population corrections seem to 
indicate that
errors up to about 0.12 mag may be possible, probably in any photometric band. 
Therefore, work
must continue to tighten the constraints on these corrections.

We also determine the mean red clump star magnitude in our LMC fields in the
J band, which could be a useful alternative to the K band should future
work reveal that population effect corrections for red clump stars in the J band
are smaller, or more reliably determined than those for the K band.
\end{abstract}


\section{Introduction}
We have recently started on an ambitious project to improve the calibrations of
a number of
stellar distance indicators which can be used to determine the distances to
nearby galaxies,
located at $\lesssim$ 10 Mpc. This project, named the ARAUCARIA project, is
particularly
focussing on the dependence of the various stellar distance indicators on galaxy
environmental
properties, like metallicity and age of the stellar populations. Among the
objects whose
usefulness for an accurate distance measurement we want to improve within the
ARAUCARIA project
are Cepheid variables, RR Lyrae stars, red clump stars, blue supergiants, and
planetary nebulae.
These stellar distance indicators span a large range in masses, typical ages,   
and evolutionary states,
and in any galaxy at least some of these object classes can be found. In the
Local Group and
other nearby galaxy groups like Sculptor, there are a number of galaxies which
contain stellar
populations of all ages (old, intermediate, young) and in which all of our
standard candles can be found
together-these are our principal target galaxies
for establishing the possible environmental dependences of our stellar standard
candles more accurately 
than hitherto done. The aims of the project have been described in more detail  
in Gieren et al. (2001),
and first results have recently been published in Pietrzy{\'n}ski et al. (2002).

As a part of the ARAUCARIA project, we are conducting a deep near-infrared (JK) 
imaging survey in several
of our target galaxies, including the Magellanic Clouds, and the Carina and
Fornax dwarf galaxies. An
immediate aim is to use red clump stars, and the tip of the red giant branch
technique to determine
the distances to these galaxies. In this paper, we are presenting a 
distance determination to the Large Magellanic Cloud from the red clump star 
method, applied in the pure infrared domain (from a
K, J-K color-magnitude diagram). The LMC is arguably the most important galaxy
in the process of
establishing an accurate extragalactic distance scale, and our current lack of  
ability to measure
extragalactic distances is best reflected in the large range of distances
measured for the LMC with
different techniques (e.g. Gibson, 2000). Regarding the red clump stars as a
distance indicator, the method was introduced in a fundamental paper of 
Paczy{\'n}ski \& Stanek (1998),
and the red clump star mean brightness was calibrated in the optical V and I 
bands in their paper.
Important
progress has recently been made by Alves (2000) who was the first to calibrate 
the absolute magnitudes
of a large sample of {\it Hipparcos}-observed nearby red clump stars in the   
K-band. An outstanding and
obvious advantage of the red clump star technique is the fact that accurate
distances were measured by
{\it Hipparcos} to
nearly a thousand of these stars in the solar neighborhood, making the red clump
star method
the currently only technique having a zero point which is set from an accurate
geometrical calibration (Paczy{\'n}ski \& Stanek 1998).   
While previous efforts to calibrate the red clump star absolute magnitude in the
optical I-band, and apply it to other galaxies were plagued by
problems with extinction corrections, and with a small, but significant
dependence of $M_{\rm I}$ on metallicity, which
manifested themselves in considerable differences between red clump distances to
the LMC using I-band magnitudes
(e.g. Udalski et al. 1998; Udalski et al. 2000; Romaniello et al. 2000),
the use of K-band magnitudes minimizes these problems, and in the case of the
LMC virtually eliminates
any dependence of the final K-band red clump distance result on the adopted
reddening. 
In  the very recent work of Grocholski \& Sarajedini
(2002) who studied population effects
on the K-band magnitude of red clump stars from 2MASS photometry of open
clusters it was suggested that $M_{K}$ may be  less dependent on stellar 
age and metallicity than $M_{I}$. Therefore, it is of a profound importance
to verify in detail possible population effects on the mean  K-band red clump 
magnitude  
in order to be able to  derive truly accurate distances with this method.

In the following, we describe what we believe is the as yet most accurate set of
near-IR data for
samples of red clump stars in 2 fields in the bar of the LMC, obtained with the
{\it New Technology
Telescope} on La Silla under superb seeing conditions, analyze these data 
together 
with optical photometry of the red clump stars in these fields taken from OGLE 
II databases,  and 
derive the distance to the LMC. We thoroughly discuss the accuracy of
our distance determination, including the effect of environmental corrections on 
our
(and other) results. In forthcoming papers, we will extend this work to a number 
of galaxies
in the Local Group, in an effort to tighten the constraints upon the appropriate
corrections which have to be applied to the observed red clump star magnitudes 
to account
for differences in the red clump star populations we are observing in these 
galaxies,
and in the solar neighborhood.

\section{Observations}
All observations presented in this paper were collected with 
the ESO/NTT telescope at the  La Silla observatory in  
Chile. The SOFI instrument in wide field mode with focal elongator 
in the grism wheel, was used. The resulting field of view was 
about 2.49 x 2.49 arcmin with a scale of 0.146 arcsec/pixel. 
This configuration guaranteed an excellent image quality
(superior to the quality achievable in other SOFI modes) with a very stable 
PSF over the whole CCD, which was a substantial advantage in deriving accurate 
PSF photometry in the dense fields of the LMC bar.

Two fields (see Table 1) were observed through Ks and J filters, 
under photometric and 0.6 arcsec seeing conditions.
In order to account for frequent sky level variations, especially 
in the K band, the observations were made with a jittering technique. In the Ks 
filter, we did 6 consecutive
10 s integrations in any given sky position before moving the telescope by about 
20 arcsec 
to a different position. We used 20 different jittering positions in Ks which 
resulted in 
a total, net exposure time of about 20 min in the Ks filter.
The J-band data were obtained in a similar fashion, but for each telescope 
jittering position
we just obtained one 10 s integration, leading to a total integration time of 
about 3 min
in this filter.
In order to accurately transform our data to the standard system, we secured 
9 observations of 7 different standard stars from the UKIRT system 
(Hawarden et al. 2000) at a variety of airmasses,
and spanning a broad range in colors bracketting the red clump star colors,
 along with our LMC fields.

\begin{deluxetable}{c c c c c}
\tablecaption{Observed fields}
%\tablewidth{0pt}
\tablehead{
\colhead{Field} & \colhead{RA (J2000)} & \colhead{DEC (J2000)} &
\colhead{$N$ stars} & \colhead{$N_{RC}$} \\
}
\startdata
FI & 05:33:39.5 & -70:09:36.5 & 1472 & 464 \\
FII& 05:33:53.8 & -70:04:01.9 & 1441 & 494 \\
\enddata
\end{deluxetable}

\section{Reductions and Calibrations}
Sky subtraction of the images was performed in a two-step process implying 
masking 
of stars with the xdimsum IRAF package. As a next step, the data were 
flatfielded 
and stacked into the final images. The PSF photometry was derived 
with the $DAOPHOT$ and $ALLSTAR$ programs. The PSF model was obtained
in an iterative way which was described in detail in Pietrzy{\'n}ski, Gieren and 
Udalski
(2002). In order to derive aperture corrections, about 50 relatively 
isolated stars were selected on each image. Then all stars 
located in the neighbourhood of those stars were iteratively subtracted, 
and  aperture photometry on the selected stars was carried out.
The median from the aperture corrections derived for all selected isolated stars 
was adopted as the final aperture correction for a given frame.
The typical rms scatter of the aperture corrections derived in this way
was about 0.008 mag.  

Aperture photometry on the standard stars was performed by choosing an aperture 
of 14 pixels. From our instrumental magnitudes, the following transformations 
to the standard system were derived:

$$J-K = 0.930*(j-k)  + 0.620 $$
$$K = k  - 0.036*(j-k) - 2.704 $$
 
where J-K and K stand for the standard  color and magnitude in the UKIRT 
system, and j-k and k denote the instrumental aperture magnitudes 
scaled to 1 s exposure times.

As can be seen, the color coefficients in the transformation equations are close 
to 1 and 0, which 
means that our instrumental system matches the standard UKIRT
system very well. The residuals did not exceed 0.02 mag and did not show 
any dependence on color or brightness (see Fig. 1).
The total error of our transformations, 
in particular on the zero point of the K and J magnitudes, was estimated to 
be less than 0.03 mag. 

In order to perform an external check on the accuracy of the zero point of our 
photometry 
we searched the 2MASS database ($ http://www.ipac.caltech.edu/2mass $) for stars 
with accurate measurements
and common to our data. Alltogether 19 such stars were found. Before comparing 
the results we need to transform our data into the system used by 2MASS 
(Carpenter
2001). After doing this, we found the following differences between our 
photometry and the 2MASS data:
$-0.01\pm0.04$ and $0.00\pm0.05$ in  K magnitude  and J-K color, respectively.
Another independent check on the zeropoint of our K-band photometry can be done 
comparing the mean brightness of the red clump stars derived in this paper (see 
next 
section) and Alves et al. (2002). Alves et al. (2002) give $<K>_{\rm KOORNNEEF} 
= 
16.974 $ mag, while we obtain $<K>_{\rm UKIRT} = 16.895 $ mag. After converting 
both values to the $(K_{\rm S})_(2MASS)$ system (Carpenter 2001) and correcting 
for the geometry of the LMC as discussed by van der Marel et al. (2002) (our 
fields 
are about 0.025 mag closer than those observed by Alves et al.), we found that 
our photometry is by about 0.001 mag brighter than that presented by Alves et 
al. 
This excellent agreement of our present K-band photometry with Alves et al. 
reassures
that systematic uncertainties on the photometric zero points are very small
in both studies.

\section{Results}
In Fig. 2  we present the observed K vs J-K color-magnitude diagrams for 
our two LMC fields, calibrated to the UKIRT system. It can be appreciated that
the faintest stars we were able to measure have K-band magnitudes of 
about 20.5. In both diagrams, the outstanding structural feature is the red 
clump, at a K magnitude
of about 17. The S/N ratio for stars at the magnitude level of the red clump 
was calculated to be around 50. To our knowledge, these are the deepest infrared
color-magnitude diagrams ever obtained in the LMC. Besides the red clump, 
many other features can be identified, including the
main sequence and red giant branch. 
A detailed discussion of the different stellar populations appearing 
in these diagrams will be subject of another study. 

In order to derive the mean observed K-band magnitudes of the red clump stars in 
both fields, all stars 
having J-K colors in the range from 0.35-0.80 mag and K-band magnitudes in the 
range from 16-18 were selected. 

The histograms of K-band magnitudes 
were derived for each field by using bins of 0.05 mag. Then the following 
function (Paczynski \& Stanek 1998):\\

$$n(K)=a+b(K-K^{\rm max})+c(K-K^{\rm max})^2+
\frac{N_{RC}}{\sigma_{\rm RC}\sqrt{2\pi}} \exp\left[-\frac{(K-K^{\rm
max})^2}{2\sigma^2_{\rm RC}}\right]\eqno(1)$$ \\

consisting of the Gaussian function component 
representing the distribution of red clump stars, superimposed on a second
order polynomial function approximating the stellar background 
 was fitted to the data. The total number of stars measured in our fields, and 
the number of stars in our fitting boxes, are indicated in Table 1.
 
In order to check on how our derived mean red clump star K-band magnitudes could 
potentially depend on 
the size of the box used for the selection of the stars for the contruction
of the histograms, we changed this box by 0.1 mag in color and 0.3 mag 
in magnitude (larger changes would be clearly unreasonable from the inspection 
of the CMDs),
and repeated the whole procedure. As a result of this exercise, we found no 
significant 
change in the derived mean magnitudes of the red clump stars, for neither field.

As a result of our fits, we derive for the mean red clump K-band magnitudes 
16.893 $\pm$ 0.011 mag
for field I, and 16.898 $\pm$ 0.010 mag for field II. 
The same procedure yields mean  J-band magnitudes of 17.498 $\pm$ 0.012
and 17.512 $\pm$ 0.011  mag, respectively, for the two fields.

For both filters, the values for fields I and II are clearly in 
very good agreement.
This is consistent with the fact that our fields are located very close to each 
other 
(the distance between them is smaller than 20 arcmin) and no differential depth 
effect due to 
the slight tilt of the
LMC bar with respect to the line of sight should be expected, at such small 
angular differences.
It is therefore justified to merge, for a given filter, 
the data of the 2 fields into one diagram in order to improve the statistical 
error on our results.
We did this, and Fig. 4 shows 
the fits of function (1) to the combined data, in both filters. From these fits, 
we derive 
as our final, adopted values for the mean K- and J-band magnitudes of red clump 
stars in 
the LMC 16.895 $\pm$ 0.007  and 17.507 $\pm$ 0.009 mag, respectively.

In order to apply appropriate extinction corrections to these observed mean 
K- and J-band magnitudes, we adopted
a value of E(B-V)=0.152 mag, corresponding to our observed fields, 
from the OGLE II map of extinction in the LMC (Udalski et al. 1999). 
These maps were determined using the mean brightness of the 
red clump stars as tracers  of the fluctuations of the mean reddening.
The zero point was adopted from three different previous studies of 
interstellar extinction based on OB stars (Udalski et al. 1998, Lee 1995), 
and colors of RR Lyrae stars (Walker 1993). As noted by Udalski et al. (1999), 
all zero points were consistent within a couple of thousands of magnitude. 
The absolute calibration of these maps was checked by comparing the
observed I-band magnitudes of the red clump stars with the extinction-free 
magnitudes
of red clump stars in clusters and field stars around them in 
the outer parts of the LMC (Udalski 1998), again showing very good agreement
(to within 0.01 mag). Therefore we adopt, as an error on the reddenings derived 
from these maps,
the value of 0.02 mag as given by Udalski. We note that from a different study 
(Harris, Zaritsky
and Thompson 1997) there is an indication that the uncertainty on the reddening 
might be
somewhat larger (about 0.035 mag), but such a slight difference would not have 
any
significant effect on the budget of systematic errors affecting our current 
distance determination of the LMC. 

Assuming the standard 
extinction curve ($A_{K} = 0.367 * E(B-V)$ and $A_{J} = 0.902 * E(B-V)$, 
Schlegel et al. 1998), 
we derive $A_{K}$ = 0.055 mag and  $A_{J}$ = 0.137
mag,  and therefore $<K>_{0}$= 16.839 and $<J>_{0}$= 17.370. 
To our knowledge, this is the first determination of the J-band mean magnitude 
of red clump stars in the LMC existing to date. 

To compare our reddening-free mean K-band magnitude with the absolute K-band 
magnitude 
of the red clump stars from the {\it Hipparcos} sample (Alves 2000), 
we adopted the transformations between the Koorneef, 2MASS and UKIRT systems as 
derived by Carpenter 
(2001). It is worthwhile noting that there is practically no difference 
between
K magnitudes in the UKIRT system to which our present data were transformed, and 
the 2MASS Ks band. 
(e.g. Carpenter (2001) gives the formula $(K_{s})_{2MASS} = K_{UKIRT} +
0.004 * (J-K)_{UKIRT} +0.002$, which results in a difference of about 0.004 mag 
for red clump stars between their $K_{UKIRT}$ and $K_{s}$ 2MASS magnitudes.).  
The correction to bring the red clump star magnitudes from the Hipparcos sample 
to the UKIRT system 
is 0.044 mag (Carpenter 2001). Applying this correction, and {\it assuming that 
there is no 
population effect} on K-band red clump star magnitudes, we obtain a LMC true 
distance modulus
of 18.487 +/- 0.008 (random) +/- 0.045 (systematic)  mag. The systematic error 
on 
this result contains contributions from
the uncertainty of the {\it Hipparcos}- calibrated absolute magnitude of red 
clump stars (0.03 mag), 
our current photometric zero point (0.03 mag), the uncertainty of the 
transformations between the different infrared 
systems (0.01 mag) and the uncertainty of the assumed reddening (0.02 mag). 

If we adopt the van der Marel et al. (2002) geometrical model of the LMC, 
the corresponding distance of the LMC barycenter is 18.501 $\pm$ 0.008 
(random) $\pm$ 0.045 mag (systematic).

\section{Discussion}

Population effects on red clump star absolute magnitudes in different 
environments, including the
solar neighborhood and the LMC, have been studied in some detail by Girardi et 
al. (1998),
and by Girardi and Salaris (2001). In particular, under the assumption of a star 
formation history 
(SFH) and chemical evolution model for the solar neighborhood and the LMC the 
latter authors 
found from population synthesis models 
that the distance moduli derived from a comparison of the mean magnitude of the
red clump stars in the LMC to the corresponding mean magnitude of the red clump 
stars 
in the Hipparcos sample should be corrected 
by -0.03, 0.2 and 0.3 mag in K, I and V, respectively, in order to compensate 
for the difference in metallicity and age  
 between these two environments. A corresponding -0.03 mag correction has been 
applied
 by Alves et al. (2002) to their K-band data to allow for the population effect. 
Applying
 the same correction, our LMC barycenter distance result becomes 18.471 mag, in
 excellent agreement with their result.
 
 Unfortunately, a discussion of
the accuracy of such corrections was not presented in Girardi and Salaris 
(2001).
An estimation of the uncertainties on these corrections
should be based on population synthesis calculations performed for very 
different SFHs and
chemical evolution models, and should also take into account current 
uncertainties on the
input physics of the stellar models used, 
which is a task far beyond the scope of this paper. However, we can make a rough 
estimate of the uncertainties 
of populations corrections using the published simulations of Girardi 
et al (1998) and Girardi and Salaris (2001) for the I band. In spite of the fact 
that they 
performed their 
simulations for a very limited range of possible SFHs and chemical evolution 
models, their data show that changing these parameters one can obtain changes in 
the synthetic mean
brightness of red clump stars  of 0.115 mag (see Table 4 in 
Girardi and Salaris 2001). This is in agreement with the error estimate of about
0.1 mag for the model calculation results which was earlier given in Girardi et 
al. (1998; their
section 5).
So far no similar data for the K band are available, 
but
it seems reasonable to assume that the current uncertainty on the population 
correction
in the K band may be of a similar order of what is extracted from the work of 
Girardi and
coworkers for the I band. It is therefore of great importance to reduce the 
current uncertainty
on population corrections for red clump stars, particularly in near-infrared 
bands, 
in order to make these objects truly superb standard candles. We note, for 
example,
that Udalski (1998, 2000) derived an I-band population correction for LMC red clump 
stars
of only 0.04 mag, based on a thorough study of red clump stars in LMC clusters, 
which is
quite different from the 0.2 mag correction derived from the models of Girardi 
and Salaris
and suggesting that there is still room for significant improvement in the 
future from both
the model, and the observational side.
 A reduction of these
uncertainties is likely to reconcile the apparent discrepancies of
the LMC distance modulus obtained in the past by different workers from data in 
different bands.
We expect that our ongoing near-infrared work on red clump stars in other Local 
Group 
galaxies which
span a range in environmental properties will lead, in the near future, to such 
an improved 
empirical calibration
of the dependence of red clump star magnitudes on the stellar ages and 
metallicities.

\acknowledgments
We are grateful to Leo Girardi for very useful and enlightening discussions 
about his red clump
star models. We are grateful to the European Southern Observatory for allocating
observing time to this project with SOFI on the NTT telescope. It is
a real pleasure to thank the NTT team for their expert support which helped to 
acquire data of
the highest possible accuracy.
 WG gratefully acknowledges financial support for this
project from the Chilean Center for Astrophysics FONDAP No. 15010003. 
He also acknowledges support
from the Centrum fuer Internationale Migration und Entwicklung in
Frankfurt/Germany
who donated the Sun Ultra 60 workstation on which our data reduction and
analysis was carried out.  Finally, we would like to thank an anonymous 
referee for  particularly interesting comments which helped to 
improve this paper.

\begin{figure}[htb]
\vspace*{10 cm}
\includegraphics{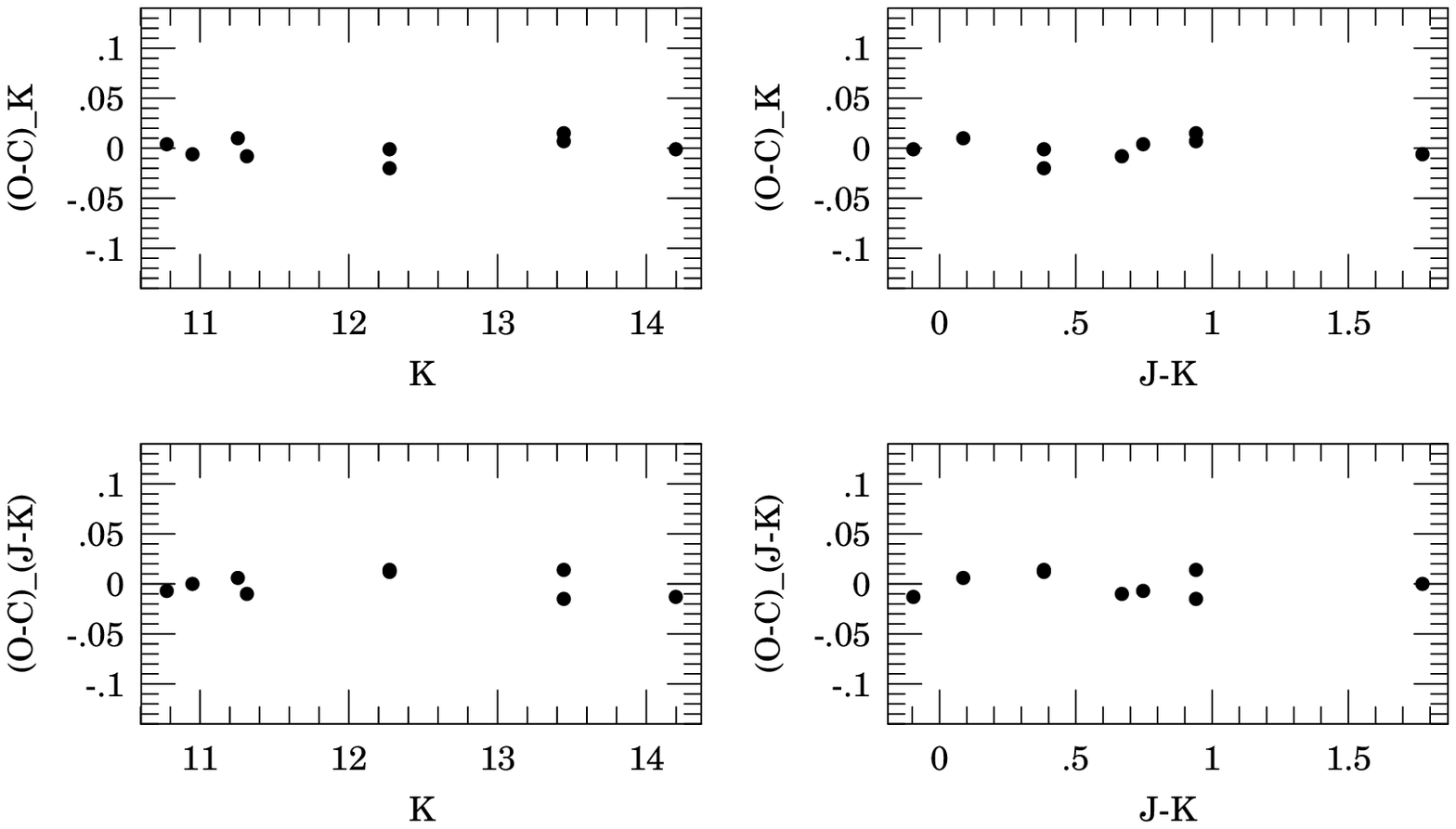}
\caption{Observed minus calculated K-band magnitudes and J-K colors for 
standard stars, as a function of magnitude and color.
}
\end{figure}

\begin{figure}[htb]
\vspace*{12 cm}
\includegraphics{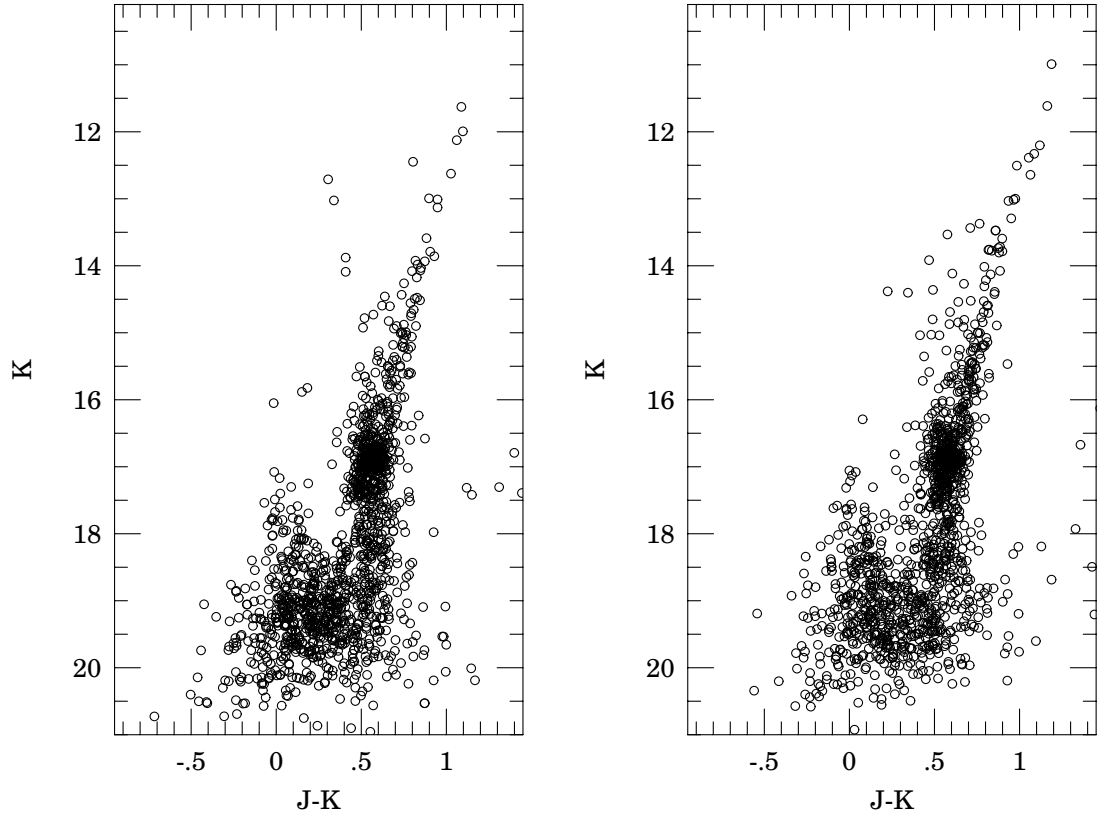}
\caption{K, J-K near-infrared color-magnitude diagram for the fields FI and 
and FII (left and right panel, respectively; see 
Table 1). The
most conspicuous feature is the red clump at K $\approx$ 17 and J-K $\approx$ 
0.6 mag. Photometry 
goes
down to about K=20.5, and the signal-to-noise ratio at the red clump magnitude 
is
about 50. The main sequence and red giant branch are also clearly delineated by 
the
data.
}
\end{figure}

\begin{figure}[htb]
\vspace*{10 cm}
\includegraphics{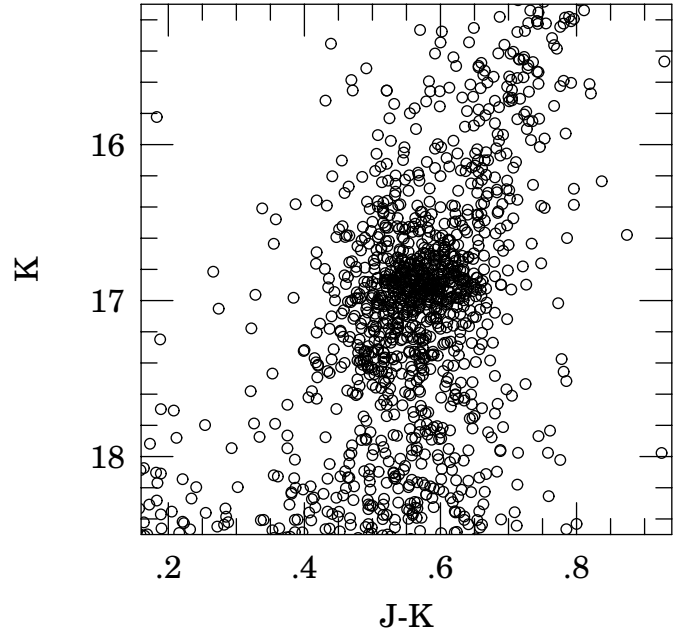} 
\caption{The combined K, J-K color-magnitude diagram for the two observed 
fields,
on a magnified scale which allows a better appreciation of the red clump 
structure.
}
\end{figure}

\begin{figure}[htb]
\vspace*{10 cm} 
\includegraphics{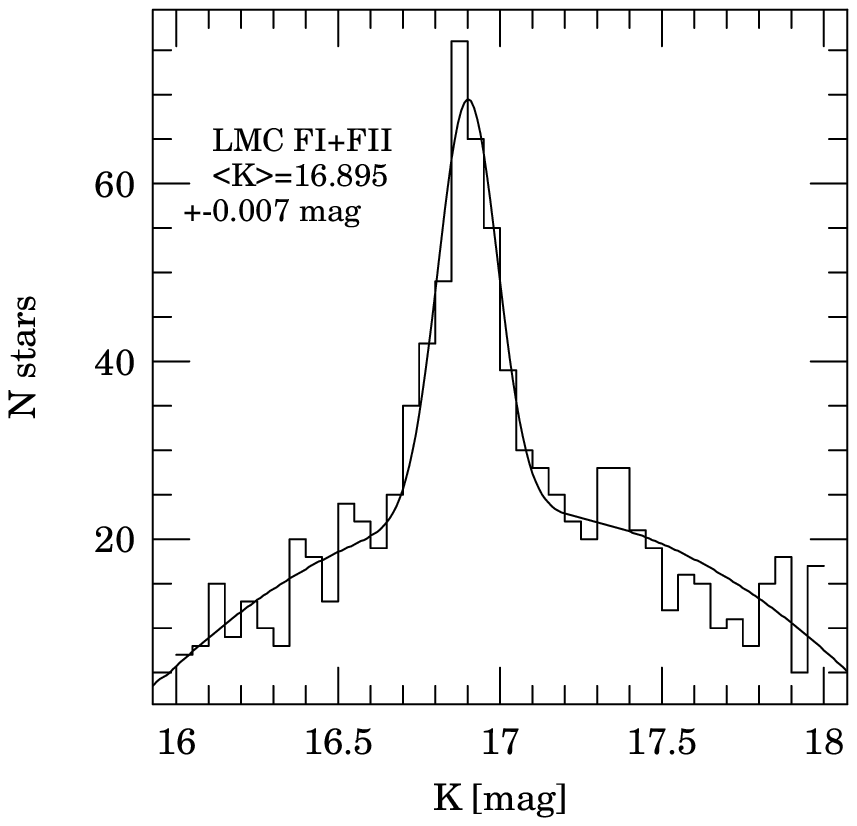}
\includegraphics{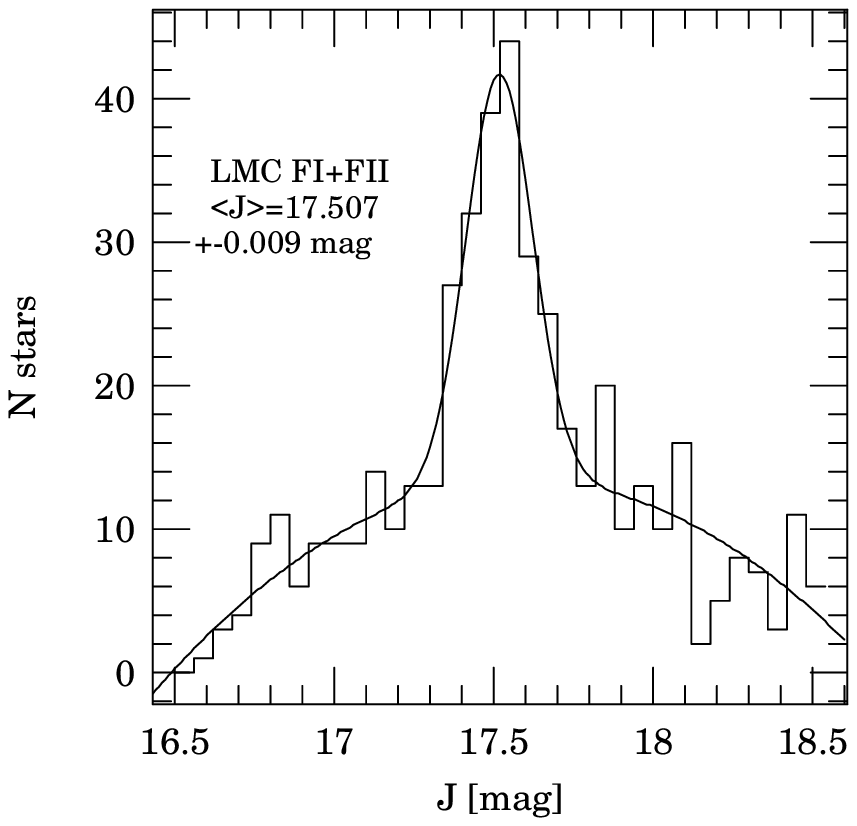} 
\caption{The Gaussian and polynomial fit, according to equation (1), applied to 
the
stars in a box around the red clump in observed fields (see text for details). 
Sharp
and well-defined peaks at a K-band magnitude of 16.895, and at a J-band 
magnitude 
of 17.507 are  obtained from the data, with excellent statistics.
}
\end{figure}

\end{document}